\documentstyle[11pt]{article}
\def\hea{I}
\def\corr{{II}}
\def\ell{\epsilon}
\def\F{{\cal F}}
\def\O{{\cal O}}
\def\AppA{A}
\def\AppB{B}
\def\AppC{C}
\def\AppD{D}
\def\D{I}
\def\g{D} 
\def\bC{\bar{C}}
\def\bg{\bar{\g}}
\def\bS{\bar{S}}
\def\bQ{\bar{Q}}
\def\by{\bar{y}}
\def\bw{\bar{w}}
\def\th{\theta} 
\def\pa{\partial}
\def\half{  {1\over 2} }
\def\tshalf{ {\textstyle {1\over 2}} }
\def\tsthird{ {\textstyle {1\over 3}} }
\def\tstwothirds{ {\textstyle {2\over 3}} }
\def\quarter{  {1\over 4} }
\def\tsquarter{ {\textstyle {1\over 4}} }
\def\cl{ {\rm classical}}
\def\oneloop{ {\rm 1-loop}}
\def\oneinst{ {\rm 1-inst.}}
\def\a{\alpha}
\def\b{\beta}
\def\G{\Gamma}
\def\d{\delta} 
\def\L{\Lambda}
\def\l{\lambda}
\def\be{\begin{equation}}
\def\ee{\end{equation}}
\def\bea{\begin{eqnarray}}
\def\eea{\end{eqnarray}}     

\renewcommand{\baselinestretch}{2.0}

\begin{document}

\begin{flushright}
BRX TH-432\\[-.2in]
BOW-PH-110\\[-.2in]
HUTP-98/A17
\end{flushright}

\begin{center}

{\Large\bf One-Instanton Test of a Seiberg--Witten Curve
from M-theory: the  Antisymmetric
Representation of SU(N) }

\vspace{.2in}

\renewcommand{\baselinestretch}{1}
\small
\normalsize
Stephen G. Naculich\\
Department of Physics\\
Bowdoin College, Brunswick, ME 04011

\vspace{.1in}

Henric Rhedin\footnote{Supported by the Swedish Natural
Science Research Council (NFR) under grant no.
F--PD1--883--305.}\\
Martin Fisher School of Physics\\
Brandeis University, Waltham, MA 02254

\vspace{.1in}

Howard J. Schnitzer\footnote{Research supported in part
by the DOE under grant DE--FG02--92ER40706.}\\
Lyman Laboratory of Physics\\
Harvard University, Cambridge, MA 02138\\
and\\
Martin Fisher School of Physics\footnote{Permanent address.\\
{\tt  \phantom{aaa} naculich@bowdoin.edu; 
rhedin,schnitzer@binah.cc.brandeis.edu}}\\
Brandeis University, Waltham, MA 02254

\vspace{.2in}

{\bf Abstract}
\end{center}

\renewcommand{\baselinestretch}{2}
\small
\normalsize
\begin{quotation}
\baselineskip14pt
One-instanton predictions are obtained from the 
Seiberg--Witten
curve derived from M-theory by Landsteiner and Lopez for the
Coulomb branch of  
$N=2$ supersymmetric 
SU(N) gauge theory
with a matter hypermultiplet in the antisymmetric representation.
Since this cubic curve describes a Riemann surface that is 
non-hyperelliptic, a systematic perturbation expansion about a 
hyperelliptic curve is developed, with a comparable expansion for 
the Seiberg--Witten
differential.  
Calculation of the period integrals of the SW differential
by the method of residues of D'Hoker, Krichever, and Phong 
enables us
to compute the prepotential explicitly to one-instanton order.

It is shown that the one-instanton predictions 
for SU(2), SU(3), and SU(4) agree with previously available results. 
For SU(N), $N\geq 5$, our analysis provides explicit predictions of
a curve derived from M-theory at the one-instanton level in field theory.
\end{quotation}

\newpage

\noindent{\bf 1. ~Introduction}

Enormous advances have been made in 
understanding the exact behavior of 
low-energy four-dimensional $N$=2 supersymmetric gauge theories
following the seminal work of Seiberg and Witten \cite{001}.
In their program one extracts the physics from a specified Riemann
surface particular to the problem, and a preferred
meromorphic 1-form, the Seiberg--Witten (SW) differential.  This
data allows one in principle to reconstruct the prepotential
of the Coulomb branch of the theory in the low-energy limit
from the period integrals of the SW differential.

For $N=2$ gauge theories based on classical groups, 
either without matter hypermultiplets 
or with matter hypermultiplets in the 
defining representation\cite{001}--\cite{010},
the associated Rieman surface is hyperelliptic.
Such theories have been studied in detail by means of  
two complementary techniques:  
the formulation and solution of the coupled set of Picard--Fuchs partial
differential equations for the periods \cite{010}--\cite{021}, 
and the direct evaluation of the
period integrals by the method of residues developed by D'Hoker,
Krichever, and Phong (DKP) \cite{022}--\cite{024}.  
The Picard--Fuchs approach has the advantage of 
being able to give global information about
the prepotential through explicit solutions to the differential equations, 
suitably analytically continued \cite{001,010,011,021}.  
However, the complexity of the set of equations 
increases rapidly with the rank of the gauge group.  
The methods of DKP \cite{022}--\cite{024}, 
on the other hand, are not severely limited
by the rank of the group, 
but results are easily obtained only for
the first few terms of the instanton expansion of the prepotential.

Not all Seiberg--Witten theories are solved by means of a
hyperelliptic surface.  String theory has provided new methods of
constructing solutions to a wide class of Seiberg--Witten problems.
In particular geometric engineering \cite{025}
and methods from M-theory \cite{026,027} have
greatly enlarged the class of $N$=2 supersymmetric gauge theories that
can be studied.  These techniques have given rise to Riemann surfaces
that are not hyperelliptic \cite{018}, and to curves that are not 
Riemann surfaces at all \cite{025}.
For these new $N$=2  theories,
the explicit formulation and solution of the appropriate Picard--Fuchs
equations may be awkward at best.  
A direct evaluation of the period integrals may be more suitable,
but systematic methods for the 
computation of the period integrals have not yet been developed for
non-hyperelliptic surfaces. 
Attention to this issue is one of the 
motivations of this paper.

Another issue that must be addressed is the {\it test} of the
curves predicted by geometric engineering or M-theory against the
results of standard $N$=2 supersymmetric theories.  
A prediction for a SW curve does not immediately translate into an
explicit expression for the prepotential.  
Although the curves
derived by string methods have been subjected to a number of 
consistency checks, 
no direct confrontation with $N$=2 field theory 
beyond checking the one-loop perturbative prepotential
has been presented. 
In particular, explicit instanton
expansions for the prepotential have not been carried out and
checked against field theory.

Using M-theory, Landsteiner and Lopez (LL) 
obtained a non-hyperelliptic curve characterizing 
the Coulomb phase of $N=2$  SU(N) gauge theory
with a single matter hypermultiplet in the 
antisymmetric or symmetric representation of the group  \cite{027}.
LL checked that the one-loop beta function of the theory had the
correct coefficient, 
that their curve had the correct limit as the mass
of the multiplet $m\rightarrow \infty$, 
and that the singular locus of 
the curve had expected singularities for SU(2) and SU(3).  
However,
instanton predictions from the LL curves are not
known.  

In this paper and its companion \cite{029}, 
we develop methods to
extract the instanton predictions
of the Landsteiner-Lopez curves \cite{027}.
We calculate the explicit one-instanton contribution to the
prepotential for $N=2$ SU(N) gauge theory  with matter 
in the antisymmetric representation.
(In ref.~\cite{029}, we do the same for the
symmetric representation.)
The key idea of this paper is the
development of a systematic perturbation scheme about a hyperelliptic
approximation to the LL curves.  
This induces a perturbative expansion for the SW differential.  
We apply the method of residues developed by DKP \cite{022}--\cite{024} 
to each term this expansion, 
which enables us to calculate the renormalized order parameters of the theory
to the one-instanton level.
{}From these, we compute the prepotential to the same order.

Our results provide a test of curves derived from M-theory
since there exists independent knowledge of the instanton expansion for
the cases of SU(2), SU(3), and SU(4).  
Specifically, 
SU(2) with matter in the antisymmetric representation
is equivalent to 
pure gauge theory \cite{001}, 
SU(3) with matter in the antisymmetric representation
is equivalent to 
SU(3) with matter in the defining representation \cite{022},
and 
SU(4) with matter in the antisymmetric representation
is equivalent to 
SO(6) with matter in the defining representation \cite{023,028}.
Happily, our results for the one-instanton predictions of the  
LL curves coincide with results previously obtained for
SU(2), SU(3), and  SO(6).  It seems to us that it is extremely important
to continue to test the field theoretic predictions of geometric 
engineering and M-theory.  This will require further developments of
the methods presented in this paper, as well as new explicit 
microscopic instanton calculations \cite{030} starting from field theory.

\noindent{\bf 2. ~The Setup}
\renewcommand{\theequation}{2.\arabic{equation}}
\setcounter{equation}{0}

The  formulation of Seiberg--Witten theory has been discussed by 
many authors, so our setup of the problem will be brief.  Consider
$N$=2 supersymmetric SU(N) gauge theory with one matter 
hypermultiplet in the antisymmetric representation.  
There is also an
$N$=2 chiral multiplet in the adjoint representation, which contains
a complex scalar field $\phi$.  This theory is asymptotically free.
Along the flat directions of the potential, $[\phi , \phi^+ ]=0$, and
the symmetry is broken to U(1)$^{N-1}$, with an $N$--1 dimensional
moduli space, parametrized classically by the eigenvalues of $\phi$
\be
e_k; \; \; 1 \leq k \leq N \; .
\ee
In terms of $N$=1 superfields, the Wilson effective Lagrangian, to
lowest order in the momentum expansion, is
\be
{\cal L}  = \frac{1}{4\pi}  Im \left[ \int d^4\th \:
\frac{\pa \F(A)}{\pa A^i} \: \bar{A}^i  
 + 
\frac{1}{2} \int d^2\th \:
\frac{\pa^2 \F(A)}{\pa A^i\pa A^j} \: W^iW^j \right]
\label{eq:22}
\ee
where the $A^i$ are $N$=1 chiral superfields.  
Holomorphy implies that prepotential $\F$ has the form
\be
\F(A)  =  \F_{\cl} (A) +  \F_{\oneloop} (A)
 	+  \sum^\infty_{d=1}  \L^{ [2N-I(R)]d} \, \F_{d-{\rm inst.}} (A) 
\label{eq:23}
\ee
where $I(R)= N-2$ is the index of the antisymmetric  representation,
and the summation is over 
instanton contributions to the prepotential.  
{}From perturbation theory one knows that 
the one-loop prepotential takes the form for massless hypermultiplets
\be
\F_{\oneloop}   =  \frac{i}{4\pi} \sum_{\a\in \Delta_+} (a\cdot\a)^2 
\log \left( \frac{a\cdot\a}{\L}\right)^2 
\; - \;
\frac{i}{8\pi} \sum_{w\in W_G} (a\cdot w)^2 
\log \left( \frac{a\cdot w}{\L} \right)^2 \label{eq:24a}
\ee
where $\a$ is summed over the positive roots $\Delta_+$ of $G$, 
$w$ are the weight vectors of the weight system $W_G$ 
corresponding to the matter representation,  
and the $a_i$  are the diagonal elements of $\phi$
rotated into the Cartan subalgebra.
This becomes
\be
\F_{\oneloop}  =   \frac{i}{8\pi}
\left[
\sum^N_{i,j=1} (a_i-a_j )^2 \log \frac{(a_i-a_j)^2}{\L^2 }
\; - \; 
\sum_{i < j} (a_i + a_j)^2 
\log
\frac{(a_i + a_j)^2}{\L^2 } \right]
\label{eq:24b}
\ee
for one massless hypermultiplet in the antisymmetric representation.

The ingredients for determining the prepotential (\ref{eq:23})
using Seiberg--Witten theory \cite{001} are 
a Riemann surface 
(which depends on the moduli) 
and a preferred meromorphic one-form $\l$, 
the Seiberg--Witten (SW) differential.  
In terms of these, one may calculate
the renormalized order parameters $a_k$ and 
their duals $a_{D,k}$ using
\be
2\pi{i} \, a_k  =  \oint_{A_{k}} \l, ~~~~~~~~~~~~~~~~~~~
2\pi{i} \, a_{D,k}  =  \oint_{B_{k}} \l 
\label{eq:25}
\ee
where $A_k$ and $B_k$ are a canonical basis of homology cycles
on the Riemann surface.
Given these, the prepotential is determined via
\be
a_{D,k}  =  \frac{\pa \F}{\pa a_k} .
\label{eq:25b}
\ee
Using arguments from M-theory, 
Landsteiner and Lopez \cite{027} proposed the curve
\be
y^3 + 2A(x)y^2 + B(x)y + L^6 = 0
\label{eq:26}
\ee
and SW differential
\be
\l = x {dy \over y},
\ee
where
\bea
L^2 & = & \L^{N+2} \nonumber \\
A(x) & = & C(x) + {\textstyle {3\over 2} } L^2 \nonumber \\ 
B(x) & = & L^2 \g(x) + 3 L^4 
\label{eq:27a}
\eea
and
\bea
C(x) & = & \tshalf x^2 \prod^N_{i=1} (x-e_i) \nonumber \\
\g(x) & = & (-1)^N \, x^2 \prod^N_{i=1} (x+e_i) \; . 
\label{eq:27}
\eea
Some properties of this curve are described in Appendix \AppB.

The purpose of this paper is to derive the one-instanton
contribution $\F_{\oneinst} (A)$ to the prepotential from the LL
curve (\ref{eq:26}).  
Since this curve is not hyperelliptic,
we have developed an extension of existing methods, 
suitable at least for the instanton expansion.  
For quantum scales $\L$ much smaller than the moduli 
(the semi-classical limit),
one conjetures that the constant term in (\ref{eq:26})
is negligible relative to the first three terms.  
The approximate equation 
\be
y^2 + 2A(x)y + B(x) \simeq 0 
\label{eq:28}
\ee
is hyperelliptic, and can be analyzed by
previously developed methods \cite{022}--\cite{024}.
This hyperelliptic approximation, however,
is not sufficiently accurate to compute the one-instanton 
contribution to the prepotential.

In Appendix \AppA, we solve (\ref{eq:26})
using a systematic perturbation expansion
about the solutions of the hyperelliptic approximation (\ref{eq:28}).
The first order correction is 
\bea
y_1 & = & 
(-A-r) -
\frac{L^6}{2Br} \: \left( A - r \right) + \ldots 
\nonumber \\
y_2 & = & 
(-A+r) +
\frac{L^6}{2Br} \: \left( A + r \right) + \ldots 
\nonumber \\
y_3 & = & - \frac{L^6}{B} + \ldots 
\label{eq:29}
\eea
where
\be
r = \sqrt{  A^2 -   B} \; .
\label{eq:210}
\ee
The approximation (\ref{eq:29}) 
is sufficiently accurate to compute 
the one-instanton contribution to $\F(A)$.  
To this order in the expansion, 
the presence of the third sheet ($y_3$) 
can be neglected, as it is not connected to the first two sheets
and has no branch cuts.

The corrections in (\ref{eq:29}) induce corrections to
the SW differential
\be
\l = \l_\hea + \l_\corr + \cdots
\label{eq:211}
\ee
where $\l_\hea$  is the usual SW differential (\ref{eq:B2}) for the
hyperelliptic curve (\ref{eq:28}), given by
\be
\l_\hea 
= {x \over A+r} d(A+r)
\simeq   {x \left( {A^\prime\over A} - {B^\prime \over 2B} \right) \over 
\sqrt{1 - {B\over A^2}}} dx
\ee
and $\l_\corr$
is the first correction to the hyperelliptic approximation.  
In Appendix \AppC, 
this correction is shown to be (\ref{eq:B3})
\bea
\l_\corr
 =   -~ { L^6 \left(A - {B\over 2A}  \right)  \over
	 B^2 \sqrt{1 - {B\over A^2}} } dx
\eea
The hyperelliptic approximation $\l_\hea$
is sufficient to obtain
$\F_{\oneloop}(A) $ for the theory, 
but the correction term $\l_\corr$ 
is necessary for the computation of 
$\F_{\oneinst}(A) $.  
Higher order corrections to the hyperelliptic approximation
do not contribute at the one-instanton level.

\noindent{\bf  3. ~ The Branch Points}
\renewcommand{\theequation}{3.\arabic{equation}}
\setcounter{equation}{0}

Before beginning the computation of the order parameters
$a_k$ and $a_{D,k}$, 
we need to locate the branch-points $x^\pm_k, \; 1 \leq k \leq N$, 
connecting sheets 1 and 2.  
(By the involution symmetry described in Appendix \AppB, 
there are also branch points connecting sheets 2 and 3 at $-x^\pm_k$, 
but these are not important in the following.)  
Setting $y_1 = y_2$, we have from (\ref{eq:29}) 
\be
0 = A^2(x^\pm_k ) - B(x^\pm_k ) + 
\frac{L^6 A(x^\pm_k)}{2B(x^\pm_k)} + \ldots \; .
\label{eq:31}
\ee
Since $B(x)$ is $\O(L^2)$, 
the last term in (\ref{eq:31})
is generally $\O(L^4)$ and therefore
not important at the one-instanton level.
For small $L$, 
the $x^\pm_k$ are close to $e_k$, 
and can be Taylor expanded
\be
x^\pm_k = e_k + \sum^\infty_{m=1} (\pm 1)^m \, L^m \, \d^{(m)}_k \;.
\label{eq:33}
\ee
Following DKP,
we introduce the residue functions  
$R_k(x)$,  $S_k(x)$,  and $S_0(x)$, defined by
\bea
\frac{3}{2C(x)} 
& = & \frac{R_k(x)}{(x-e_k)} ,\nonumber \\ [.1in]
\frac{\g(x)}{C^2(x)} &=& \frac{S_k(x)}{(x-e_k)^2} = \frac{S_0(x)}{x^2} .
\label{eq:34a}
\eea
These may be written explicitly
\bea
R_k(x) &=& \frac{3}{ x^2\prod_{i\neq k} (x-e_i) } 
\nonumber \\[.1in]
S_k(x) &=& \frac{4(-1)^N\prod^N_{i=1} (x+e_i)} 
{x^2\prod_{i\neq k} (x-e_i)^2}
\nonumber\\[.1in]
S_0(x) &=&
\frac{4(-1)^N \prod^N_{i=1} (x+e_i) }
{\prod_i (x-e_i)^2} \; .
\label{eq:34b}
\eea
With these definitions 
the $\d_k^{(m)}$ may be computed from (\ref{eq:31}) to be
 \bea
\d^{(1)}_k & = & S_k (e_k)^{\frac{1}{2}} \nonumber \\ [.1in]
\d^{(2)}_k & = & \frac{1}{2} \, \frac{\pa S_k}{\pa x} (e_k)
- R_k (e_k) \; .
\label{eq:34}
\eea

We end this section by deriving an identity needed below.
To the accuracy required, we may write
$A^2(x^-_k) = B(x^-_k )$, that is
\be
{1\over 4} (x^-_k)^2 \prod_i (x^-_k - e_i)^2 
\left[ 1 + \frac{3L^2}{2C(x^-_k)} \right]^2
 =  \, L^2 (-1)^{N+2} \prod_i (x^-_k + e_i) 
\left[ 1 + \frac{3L^2}{\g(x^-_k)} \right] \; .
\label{eq:36}
\ee
Taking the logarithm of both sides of this equation and expanding, 
we obtain
\bea
0 & = & 
	2 \log 2 - 2 \log x^-_k - 2 \sum_i \log (x^-_k - e_i) 
 -  \frac{3L^2}{C(x^-_k)} + \frac{9L^4}{4C^2(x^-_k)}  \nonumber \\[.1in]
&& + 2 \log L 
 +  (N+2) \log (-1) + \sum_i \log (x^-_k + e_i ) 
 +  \frac{3L^2}{\g(x^-_k)} 
\label{eq:37}
\eea
where we need keep the $L^4$ term since
$C(x_k^-)$ contains a factor of $(x_k^- - e_k)$,
which by  (\ref{eq:33})
is $\O(L)$.

\noindent{\bf 4. ~The Order Parameters}
\renewcommand{\theequation}{4.\arabic{equation}}
\setcounter{equation}{0}

A canonical homology basis for a hyperelliptic curve
is described by DKP  \cite{022}.
The basis $A_k, \; B_k, \; 2 \leq k \leq N$ is obtained by choosing $A_k$
to be a simple contour enclosing the slit from $x^-_k$ to $x^+_k$, 
and $B_k$ to consist of the curves going from 
$x^-_1$ to $x^-_k$ on the first sheet and from
$x^-_k$ to $x^-_1$ on the second.

The renormalized order parameters $a_k$ are given by 
\bea
2\pi{i} \, a_k & = & \oint_{A_k} \l \nonumber \\
	& \approx & \oint_{A_k} \l_\hea + \l_\corr \nonumber \\
& = &
\oint_{A_k} dx \left[
\frac{x \left( \frac{A^\prime}{A} - \frac{B^\prime}{2B} \right)}
{\sqrt{ 1-\, \frac{B}{A^2}}} 
 - L^6 { \left(A - {B\over 2A}  \right)  \over
	 B^2 \sqrt{1 - {B\over A^2}} }
\right]
\label{eq:310}
\eea
The second term in (\ref{eq:310}) makes
no contribution to $a_k$ to $\O(L^2)$, 
as it has no poles at $x = e_k$ to that order.  
The first term in (\ref{eq:310}) is identical 
to what one would obtain for a hyperelliptic curve.
A residue calculation essentially
identical to eqs. (3.2)-(3.10) of ref.\cite{022} yields 
\be
a_k = e_k + L^2 \left[ \frac{1}{4} \frac{\pa S_k}{\pa x} (e_k)
- R_k (e_k) \right] + \cdots
\label{eq:311}
\ee

\noindent{\bf 5. ~The Dual Order Parameters}
\renewcommand{\theequation}{5.\arabic{equation}}
\setcounter{equation}{0}

The dual order parameters are given by
\be
2\pi{i} \, a_{D,k} 
=   \oint_{B_k} \l 
=   \oint_{B_k} \l_\hea + \l_\corr
\label{eq:41}
\ee
One evaluates the SW differential  for the $B_k$
cycle by means of a contour that goes from $x^-_1$ to $x^-_k$ on
sheet 1, crosses the branch cut at $e_k$ to sheet 2, 
runs back from $x^-_k$  to $x^-_1$ on sheet 2, and 
passes back to sheet 1 through the branch cut at $e_1$.  
{}From the results of Appendix \AppC, 
both $\l_\hea$ and $\l_\corr$  on sheet 2 differ only by a
sign from the corresponding SW differentials on sheet 1,
so we have
\bea
2\pi{i} \, a_{D,k} 
& = & 2 \int^{x^-_k}_{x^-_1} dx  (\l_\hea + \l_\corr) \nonumber \\
& = & 2 \int^{x^-_k}_{x^-_1} dx  
\left[
\frac{x \left( \frac{A^\prime}{A} - \frac{B^\prime}{2B} \right)}
{\sqrt{ 1-\, \frac{B}{A^2}}} 
 - L^6 { \left(A - {B\over 2A}  \right)  \over
	 B^2 \sqrt{1 - {B\over A^2}} }
\right]
\label{eq:41a}
\eea

\noindent {\bf (a) Hyperelliptic Approximation}

The dual order parameter in the hyperelliptic approximation is
given by 
\bea
(2\pi i \: a_{D,k} )_{\hea}
& = & \oint_{B_{k}} \l_\hea
			 \nonumber\\
& = & \lim_{\xi \to 1}  \int^{x^-_k}_{x^-_1} dx \:
\frac{2x \left( \frac{A^\prime}{A} - \frac{1}{2} \, 
\frac {B^\prime}{B} \right)}
{ \sqrt {1-\xi^2 \, \frac{B}{A^2} } }
\label{eq:42}
\eea
Following DKP \cite{022},
we have introduced a complex parameter $\xi$ with $|\xi| < 1$
so that the denominator can be expanded in a power series in $\xi^2$,
\be
(2\pi i \: a_{D,k} )_{\hea}
 \equiv  \sum_{m=0}^\infty  \D_m, 
\label{eq:43a}
\ee
where
\be
\D_m  =  
\frac{2 \xi^{2m}\G  \left( m+ \frac{1}{2} \right) }
{\G  \left( \frac{1}{2} \right) \G (m+1)} \;
\, \int^{x^-_k}_{x^-_1} dx \  x \,
\left( \frac{A^\prime}{A} -  \frac{B^\prime}{2B} \right)
\left( \frac{B}{A^2} \right)^m \; .
\label{eq:43}
\ee
When all the terms have been calculated and resummed, 
we will set $\xi\to 1$.

We begin by computing 
\be
\D_0 = 2 \, \int^{x^-_k}_{x^-_1} dx \,x
\left( \frac{A^\prime}{A} -  \frac{B^\prime}{2B} \right)
\; . 
\label{eq:44}
\ee
{}Using (\ref{eq:27a}), we expand the integrand in powers of $L$
\bea
\frac{A^\prime}{A} 
& = & 
\frac{C^\prime}{C}  +
\frac{d}{dx} \: 
\left( \frac{3L^2}{2C} - \frac{9L^4}{8C^2} + \cdots \right) \nonumber\\ [.1in]
\frac{B^\prime}{B} 
& = & 
\frac{\g^\prime}{\g} + 
\frac{d}{dx} \: 
\left( \frac{3L^2}{\g } + \cdots \right) 
\label{eq:47}
\eea
keeping only those terms that will contribute at the 
1-instanton level, $\O(L^2)$.
(We kept the $\O(L^4)$ term in the first expression 
because $C(x)$ is $\O(L)$ when $x \to x_k^-$.)
Inserting (\ref{eq:47}) into (\ref{eq:44}) and integrating by parts,
we have to that order
\be
\D_0 = \D_{0a} + \D_{0b} + \D_{0c}
\ee
where
\bea
\D_{0a} 
& = & 
2 \int^{x^-_k}_{x^-_1} dx \, x 
\left[ \frac{C^\prime}{C} - \frac{\g^\prime}{2\g} \right] 
\label{eq:48} \\[.2in]
\D_{0b} 
& = & 
 \left. 
	\left[ \frac{3xL^2}{C} 
          - \frac{9xL^4}{4C^2} 
	  - \frac{3xL^2}{\g} \right] 
\right|^{x^-_k}_{x^-_1}\label{eq:49} \\[.2in]
\D_{0c} 
& = & 
 \int ^{x^-_k}_{x^-_1} dx
\left[-\, \frac{3L^2}{C} 
          + \frac{9L^4}{4C^2} 
	  + \frac{3L^2}{\g} \right].
\label{eq:410}
\eea
{}From (\ref{eq:27}), we have
\bea
\frac{C^\prime}{C}  &=& \frac{2}{x} + \sum^N_{i=1} \, \frac{1}{x-e_i}
	\nonumber \\ [.1in]
\frac{\g^\prime}{\g} &=& \frac{2}{x} + \sum^N_{i=1} \, \frac{1}{x+e_i}
\label{eq:412}
\eea
from which we obtain
\be
\D_{0a} =  (N+2) x^-_k + 2 \sum^N_{i=1} \,
      e_i  \log  (x^-_k - e_i) 
 + \sum^N_{i=1} \, e_i  \log  (x^-_k + e_i) 
\label{eq:413}
\ee
where here, and throughout this section, 
we will explicitly write {\it only} the contribution from the
upper limit of integration.
The contribution from the lower limit of integration will be
identical, but with $k$ replaced by $1$.
Combining  (\ref{eq:413}) and (\ref{eq:49}),
and adding the product of $x_k^-$ with (\ref{eq:37}) to the result,
we find
\bea
\D_{0a} + \D_{0b} & = &
[N + 2 + 2 \log 2 + (N+2) \log (-1) +  2 \log L  ] x^-_k 
\label{eq:417a}
\\ [.1in]
&& -  2 \sum_{i} (x_k^- - e_i) \log (x_k^- -e_i) +
        \sum_i (x_k^- + e_i)  \log (x_k^-+e_i) 
        -  2 x_k^- \log x_k^- 
\nonumber 
\eea
Using eqs. (\ref{eq:33}),(\ref{eq:34b}),  and  (\ref{eq:34}), we obtain
\bea
\D_{0a} + \D_{0b} & = &
2 x_k^-
+ [N  + 2 \log 2 + (N+2) \log (-1) + 2 \log L] e_k  \nonumber \\ [.1in]
&& -  2 \sum_{i\neq k} (e_k-e_i) \log (e_k-e_i) +
      \sum_i (e_k+e_i)  \log (e_k+e_i) -  2 e_k \log e_k \nonumber \\ [.1in]
&& -  \frac{1}{2} \, L^2 \frac{\pa S_k}{\pa x} (e_k)
+ 2 L^2 R_k (e_k).
\label{eq:418}
\eea
Turning now to the term $\D_{0c}$ (\ref{eq:410}),
the middle integral is proportional to $L^3$,
which can be neglected to 1-instanton accuracy.  
Using the identities
\bea
\frac{1}{x^2 \prod_i (x-e_i)}  
&=& 
\sum_j  \frac{1}{e_j^2 \prod_{i\neq j} (e_j-e_i)} \left[ \frac{1}{ x-e_j} 
 -  \frac{1}{x} - \frac{e_j}{x^2} \right]
\nonumber \\[.2in]
\frac{1}{x^2 \prod_i (x+e_i)} 
& =&  
\sum_j \,
\frac{(-1)^N}{e_j^2 \prod_{i\neq j} (e_j-e_i)}
\left[ -\;  \frac{1}{x+e_j} +\frac{1}{x} - \frac{e_j}{x^2} \right]
\label{eq:421}
\eea
the first and third terms yield
\bea
\D_{0c} 
& = &  
\sum_{j} 
\frac{L^2}{e_j^2 \prod_{i\neq j} (e_j-e_i)}
\left[-6 \log  \left( x_k^- -e_j \right)
+ 9 \log e_k - 3  \log  \left( e_k+e_j \right) - 3\frac{e_j}{e_k} \right] 
	\nonumber\\ [.2in]
& = & 
L^2 \sum_{j}  R_j (e_j)
\left[ -2 \log  \left( x_k^- -e_j \right)
+ 3 \log e_k -  \log  \left( e_k+e_j \right)  -  \frac{e_j}{e_k} \right]  
\label{eq:422} \\ [.2in]
& = & 
- {3\over 4} L^2 \frac{\pa S_0}{\pa x} (0) \log e_k
+ {3\over 4} L^2 \frac{S_0(0)}{e_k}
- L^2 \sum_{j}  R_j (e_j)
\left[ 2 \log  \left( x_k^- -e_j \right) 
+  \log  \left( e_k+e_j \right)  \right] 
\nonumber
\eea
where we have used $x_k^- = e_k + \O(L)$, (\ref{eq:33})  
and the identities (\ref{eq:C1}) and (\ref{eq:C2}) derived in Appendix \AppD.

Next we compute the $m \geq 1$ terms in the series (\ref{eq:43a}).  
Using the identity  \cite{022}
\be
x \left( {A^\prime \over A} - {B^\prime \over 2 B} \right) 
	\left( B \over A^2\right)^m
= - {d\over dx} \left[ {x \over 2m} \left(B\over A^2\right)^m \right]
	+ {1\over 2m} \left( B \over A^2\right)^m
\label{eq:423}
\ee
together with the result (\ref{eq:31})
\be
\frac{B(x_k^-)}{A^2(x_k^-)} = 1  + \O(L^3)
\label{eq:424}
\ee
we obtain
\be
\D_m = 
\frac{\xi^{2m} \G\left( m+\frac{1}{2} \right) }
{m \G \left( \frac{1}{2} \right) \G (m+1) } 
\left[ - x^-_k + \int^{x^-_k}_{x^-_1} dx
\left( \frac{B}{A^2} \right)^m \; \right]
\label{eq:425}
\ee
(suppressing as usual the contribution from the
lower limit of integration).
One may expand
\bea
\left( B \over A^2 \right)^m 
& = &
\left( L^2 \g \over C^2 \right)^m 
\left( 1 + {3 L^2 \over \g}  \right)^m 
\left( 1 + {3 L^2 \over 2 C} \right)^{-2m}
\nonumber\\ [.1in]
& = & \sum^m_{r=0} \sum^\infty_{n=0} \G_{m,r} \tilde{\G}_{m,n}
L^{2m+2n+2r}  \left( \frac{\g^{m-r}}{C^{2m+n}} \right) \; .
\label{eq:429}
\eea
where
\bea
\G_{m,r} 
& = & \frac{3^r \G(m+1)}{\G (m-r+1)\G (r+1)}
\nonumber\\ [.1in]
\tilde{\G}_{m,n} 
& = &\frac{ (-3/2)^n \G (2m+n)}{\G (2m)\G (n+1)}
\label{eq:429b}
\eea
Higher powers of $L$ must be retained in the expansion, 
as the integration in (\ref{eq:425}) produces negative powers of $L$.
Next expand the terms in (\ref{eq:429})  in partial fractions
\be
\frac{\g^{m-r} (x)}{C^{2m+n}(x)}  = 
\sum^{2m+2n+2r}_{p=1} 
\frac{Q^{(2m,2n,2r)}_{0,p}}{x^p}
+ \sum^N_{i=1} \sum^{2m+n}_{p=1} 
\frac{Q^{(2m,2n,2r)}_{i,p}} {(x-e_i)^p}.
\label{eq:430}
\ee
We may split
\be
\int^{x^-_k}_{x^-_1} dx \left( \frac{B}{A^2} \right)^m 
= \int^{x^-_k}_{x^-_1} dx \left( \frac{B}{A^2} \right)^m_{p=1} 
+\int^{x^-_k}_{x^-_1} dx \left( \frac{B}{A^2} \right)^m_{p>1} 
\label{eq:430b}
\ee
treating the $p=1$ and $p>1$ terms in the partial fraction expansion
(\ref{eq:430}) separately.  
First
\be
\int^{x^-_k}_{x^-_1} dx \left( \frac{B}{A^2} \right)^m_{p=1}
 = \sum^m_{r=0} \sum^\infty_{n=0} \G_{m,r} \tilde{\G}_{m,n}
L^{2m+2n+2r} 
 \left[ 
Q^{(2m,2n,2r)}_{0,1} \log  x^-_k 
+
\sum^N_{i=1} Q^{(2m,2n,2r)}_{i,1} \log (x^-_k-e_i) 
\right] \; . 
\label{eq:431}
\ee
Since $\log x^-_k = \log e_k + \O(L)$, 
only the $m=1$, $n=0$, $r=0$ coefficient for the first term in (\ref{eq:431})
need be evaluated 
\be
Q^{(2,0,0)}_{0,1}
=
\frac{1}{2\pi i} \oint_{x=0} dx \, \frac{\g}{C^2} 
 = 
\frac{1}{2\pi i} \oint_{x=0} dx \, \frac{S_0(x)}{x^2} 
 = 
\frac{\pa S_0}{\pa x} (0) \; .
\ee
The coefficients of the second term in (\ref{eq:431}) are
\be
Q^{(2m,2n,2r)}_{i,1} 
= \frac{1}{2\pi{i}} \,
\oint_{A_i} dx \, \frac{\g^{m-r}}{C^{2m+n}}
\label{eq:433}
\ee
allowing us to resum the series to obtain
\be
\int^{x^-_k}_{x^-_1} dx
\left(  \frac{B}{A^2} \right)^m_{p=1}  = 
L^2 \d_{m,1} \frac{\pa S_0}{\pa x} (0) \, \log e_k \; 
+
\frac{1}{2\pi i} \sum^N_{i=1}  \log
(x^-_k - e_i )  \oint_{A_i} dx 
\left(  \frac{B}{A^2} \right)^m .
\label{eq:434}
\ee
Next consider the $p>1$ contribution to (\ref{eq:430b}). 
Integration of (\ref{eq:429})--(\ref{eq:430}) yields
\bea
\int^{x^-_k}_{x^-_1} dx
\left(  \frac{B}{A^2} \right)^m_{p>1}
&=&  \sum^m_{r=0} \sum^\infty_{n=0} \G_{m,r}
\tilde{\G}_{m,n} \: L^{2m+2n+2r}  \times
		\nonumber \\[.1in]
&& \left[
\sum^{2m+2n+2r}_{p=2}
\frac{Q^{(2m,2n,2r)}_{0,p}}
{(1-p)(x^-_k )^{p-1}}  
   + \sum^N_{i=1} \sum^{2m+n}_{p=2}
\frac{Q^{(2m,2n,2r)}_{i,p}}
{(1-p)(x^-_k - e_i)^{p-1}}  \right].
\label{eq:435}
\eea
Except for the $i=k$ term in the sum,
one only need keep the $m=1$, $n=0$, $r=0$ term  
to order $L^2$. 
For the $i=k$ terms
one needs {\it all} $m$
due to the factors of $(x_k^- - e_k)$ in the denominator,
but only $r=0$ and $n=0,1$ 
to obtain terms of
${\O}(L^2)$.  
Therefore
\bea
\int^{x^-_k}_{x^-_1} dx \left( \frac{B}{A^2} \right)^m_{p>1}
& = &  - L^2 \, \d_{m,1} 
\left[ 
\frac{Q^{(2,0,0)}_{0,2}} {x^-_k} 
+ 
\sum_{i\neq k} \, \frac{Q^{(2,0,0)}_{i,2}} {(x^-_k - e_i)} 
\right] \nonumber \\[.1in]
&& +  L^{2m}
\left[ \frac{Q^{(2m,0,0)}_{k,2m}}
{(1-2m)(x^-_k - e_k)^{2m-1}} + 
\frac{ \th_{m-2}Q^{(2m,0,0)}_{k,2m-1}}
{(2-2m)(x^-_k - e_k)^{2m-2}}
\right] 
\label{eq:437}
\\[.1in]
&& - 3m L^{2m+2}
\left[ \frac{ Q^{(2m,2,0)}_{k,2m+1}}
{(-2m)(x^-_k - e_k)^{2m}} \right] \nonumber
\eea
where $\th_{s} = 1$ for $s \geq 0$ and 
$\th_{s} = 0$ for $s < 0$.
{}From (\ref{eq:33}), we have
\be
\frac{1}{(x^-_k - e_k)^s} =
\frac{(-1)^s}{L^s \left(\d^{(1)}_k\right)^s}
\left[ 1 + sL \, \frac{\d^{(2)}_{k}}{\d^{(1)}_{k}} + \ldots \right]
\label{eq:438}
\ee
which can be used to simplify (\ref{eq:437}).  
Combining this result with (\ref{eq:425}) and (\ref{eq:434}), 
one finds 
\bea
\D_m    = 
\frac{\xi^{2m} \G\left( m+\frac{1}{2} \right) }
{m \G \left( \frac{1}{2} \right) \G (m+1) } 
\Bigg[
&-& x^-_k 
+ \sum^N_{i=1} \log (x^-_k - e_i) \frac{1}{2\pi{i}}
\oint_{A_i} dx \left( \frac{B}{A^2} \right)^m \nonumber \\[.1in]
& + & L^2 \d_{m,1} \frac{\pa S_0}{\pa x} (0) \: \log \, e_k 
 -   L^2 \, \d_{m,1} \left( 
\frac{Q^{(2,0,0)}_{0,2}}{e_k}
+ 
\sum_{i\neq k} \,  \frac{Q^{(2,0,0)}_{i,2}}{e_k - e_i} 
\right)
\nonumber \\ [.2in]
& + & L \: \frac{    Q^{(2m,0,0)}_{k,2m} } {(2m-1) \left( \d^{(1)}_k \right)^{2m-1} } 
 +  L^2  \: \frac{   Q^{(2m,0,0)}_{k,2m}  \d_k^{(2)} }
            { \left( \d^{(1)}_k \right)^{2m}} 
\label{eq:439} \\[.2in]
& + &
L^2 \, \th_{m-2}
\frac{Q^{(2m,0,0)}_{k,2m-1}} {(2-2m) \left( \d^{(1)}_k \right)^{2m-2}} 
 + 
\frac{3}{2} \, L^2 \:
\frac{Q^{(2m,2,0)}_{k,2m+1}}
{\left( \d^{(1)}_k \right)^{2m}} \Bigg] \; . \nonumber
\eea
The coefficients of the partial fraction expansion may be evaluated by
comparing (\ref{eq:430}) with (\ref{eq:34a}) and (\ref{eq:34b})
to obtain
\bea
Q^{(2m,0,0)}_{k,2m} 
& = & 
S_k(e_k)^m        
	\nonumber \\ [.1in]
Q^{(2,0,0)}_{0,2} 
&=& 
S_0(0)  
	\nonumber \\ [.1in]
Q^{(2m,0,0)}_{k,2m-1} 
& = &
m \: S_k(e_k)^{m-1}  \frac{\pa S_k}{\pa x} (e_k) 
	\nonumber \\ [.1in]
Q^{(2m,2,0)}_{k,2m+1} 
&=&
 {\textstyle{2 \over 3}} S_k(e_k)^m R_k (e_k)
\label{eq:445}
\eea
{}Using  (\ref{eq:445}) and  (\ref{eq:34}),  
equation (\ref{eq:439}) becomes
\bea
\D_m & = &
\frac{\xi^{2m} \G\left( m+\frac{1}{2} \right) }
{m \G \left( \frac{1}{2} \right) \G (m+1) } 
\Bigg[ -x^-_k + \sum _i  \log  (x^-_k - e_i) \frac{1}{2\pi i}
\oint_{A_i} dx \left( \frac{B}{A^2} \right)^m \nonumber \\[.1in]
&& + L^2 \, \d_{m,1}  \frac{\pa S_0(0)}{\pa x} \: \log \, e_k 
 -   L^2 \, \d_{m,1} 
\left( 
\frac{S_0(0) }{e_k} 
+
\sum_{i\neq k} \frac{S_i (e_i) }{e_k - e_i} 
\right)
\nonumber \\ [.2in]
&& + L \: \frac{S_k(e_k)^{\frac{1}{2}}}
{(2m-1)} 
 +  {1\over 2} L^2 \frac{\pa S_k}{\pa x} (e_k) 
 +  L^2 \th_{m-2} \left( \frac{m}{2-2m} \right)
\frac{\pa S_k}{\pa x} (e_{k}) \Bigg] \; 
\label{eq:446}
\eea
our final expression for $\D_m$, $m\geq 1$.

To carry out the sum of (\ref{eq:446}) over $m$,
several identities found in Appendix A of ref. \cite{022} 
are useful
\bea
&&\sum^\infty_{m=1} \:
\frac{\G\left( m+\frac{1}{2}\right) }
{\G\left(\frac{1}{2}\right) \G(m+1)}\:
\frac{1}{2m(2m-1)} = 1 - \log 2  
\label{eq:447}\\[.2in]
&&\sum^\infty_{m=2} \:
\frac{\G\left( m+\frac{1}{2}\right) }
{\G\left(\frac{1}{2}\right) \G(m+1)}\:
\frac{1}{2m(2m-2)} = -\tsquarter \log  2 + \tsquarter
\label{eq:448}\\[.2in]
&&\sum^\infty_{m=1} \:
\frac{\G\left( m+\frac{1}{2}\right) }
{\G\left(\frac{1}{2}\right) \G(m+1)m}\:
= 2 \log  2 \; .
\label{eq:449}
\eea
First, we separately sum the second term in (\ref{eq:446}),
using the identity (\ref{eq:423}) 
and the fact that a total derivative does not contribute to the $A$-cycle 
\bea 
\frac{1}{2\pi i} \oint_{A_i} \, &dx&\sum^\infty_{m=1} \,
\frac{\G\left( m+\frac{1}{2}\right)\xi^{2m}}
{\G\left(\frac{1}{2}\right) \G(m+1)}
\: \frac{1}{m} \:
\left( \frac{B}{A^2} \right)^m
\nonumber\\[.1in]
& = &\frac{1}{2\pi i} \oint_{A_i} \, dx 
\  2x \left( {A^\prime \over A} - {B^\prime \over 2B} \right)
\sum^\infty_{m=1} \,
\frac{\G\left( m+\frac{1}{2}\right)\xi^{2m}}
{\G\left(\frac{1}{2}\right) \G(m+1)}
\left( \frac{B}{A^2} \right)^m
\nonumber\\[.1in]
& = & \; \frac{1}{2\pi i} \oint_{A_i} \, dx \ 2x
\left( \frac{A^\prime}{A} - \frac{B^\prime}{2B} \right)
\left[ \frac{1}{\sqrt{1-\xi^2 \, \frac{B}{A^2}}} - 1 \right]
\nonumber\\[.1in]
& = &
{2 \over 2 \pi i} \oint_{A_i} \l_\hea
- {2 \over 2 \pi i} \oint_{A_i} x 
\left( 
\frac{C^\prime}{C} + \frac{d}{dx} \: \frac{3L^2}{2C} + \cdots 
\right) \nonumber \\[.1in]
& = & 2  a_i - 2  e_i + 2 L^2 R_i(e_i) .
\eea
The full sum of (\ref{eq:446}) over all $m \ge 1$ is then
\bea
\sum^\infty_{m=1} \D_m & = & -
(2 \log 2) x^-_k  
+  2 \sum_i \left[ a_i - e_i + L^2 R_i (e_i) \right] \log (x^-_k - e_i ) 
\nonumber \\[.2in]
&& + {1\over 2} L^2 \frac{\pa S_0}{\pa x} (0) \log e_k  
 -  \frac{1}{2} \, L^2 \left( 
\frac{S_0(0)}{e_k} +
\sum_{i\neq k} \frac{S_i(e_i) }{e_k-e_i}  
\right) 
\nonumber\\[.2in]
&& + (2-2 \log 2) L \: S_k(e_k)^{\frac{1}{2}} 
 +  \left( \tshalf \log 2 - \tsquarter \right)
L^2 \frac{\pa S_k}{\pa x} (e_k) \; . 
\label{eq:451}
\eea

We now assemble the contributions 
(\ref{eq:418}),
(\ref{eq:422}),
and
(\ref{eq:451})
to the hyperelliptic approximation of the dual order parameter
and use 
\bea
2 (a_k - e_k) \log (x_k^- - e_k)
& =& (a_k - e_k)\left[  2\log L + \log S_k(e_k)  \right] \nonumber\\ [.1in]
& =& (a_k - e_k)\bigg[ 2\log L + 2 \log 2 + (N+2) \log (-1)   \nonumber\\ [.1in]
& & + \sum_{i} \log (e_k+e_i) 
- 2 \log e_k -2 \sum_{i\neq k} \log (e_k-e_i) \bigg]
\label{eq:452a}
\eea
to obtain
\bea
(2 \pi i a_{D,k})_\hea
& = &
(\D_{0a} + \D_{0b})
+ (\D_{0c} )
+ (\sum^\infty_{m=1} \D_m) 
				\label{eq:452} \\ [.1in]
& = & 
(2-2\log2) x_k^- + [N  + 2 \log 2 + (N+2) \log (-1) + 2 \log L] a_k  
                  + [-2 - N] (a_k-e_k)
				\nonumber \\ [.1in]
&& - 2 \sum_{i\neq k} (a_k-a_i) \log (e_k-e_i) +
      \sum_i (a_k+e_i)  \log (e_k+e_i) -  2 a_k \log e_k 
				\nonumber \\ [.1in]
&& - \tsquarter  L^2 \frac{\pa S_0}{\pa x} (0) \log e_k
	+ \tsquarter L^2 \frac{S_0(0)}{e_k}
	- L^2 \sum_{j}  R_j (e_j) \log  \left( e_k+e_j \right)  
				\nonumber\\ [.1in]
&& - \tshalf \, L^2 \sum_{i\neq k} \frac{S_i(e_i) }{e_k-e_i}  
 	+  (2-2 \log 2) L \: S_k(e_k)^{\frac{1}{2}} 
 	+  \left( \tshalf \log 2 - \tsquarter \right)
L^2 \frac{\pa S_k}{\pa x} (e_k) \; . 
				\nonumber
\eea
Using (\ref{eq:33}), (\ref{eq:34}), and (\ref{eq:311}) 
to rewrite (\ref{eq:452})
completely in terms of $a_k$, 
and using the identities (\ref{eq:C3})-(\ref{eq:C5}),
we finally obtain the result, accurate to $\O(L^2)$
\bea
(2 \pi i a_{D,k})_\hea
& = &
[N  + 2 + (N+2) \log (-1) + 2 \log L] a_k  
				\nonumber \\ [.1in]
&& - 2 \sum_{i\neq k} (a_k-a_i) \log (a_k-a_i) +
      \sum_i (a_k+a_i)  \log (a_k+a_i) -  2 a_k \log a_k 
				\nonumber \\ [.1in]
&& -   \tsquarter L^2 \frac{\pa S_0}{\pa x} (0) \log a_k
	- \tsquarter L^2 \sum_{j}  
	\frac{\pa S_j}{\pa x} (a_j) \log  \left( a_k+a_j \right)  
\nonumber \\ [.1in]
&& +    \tsquarter L^2 \frac{\pa S_k}{\pa x} (a_k) \;
+ \tsquarter L^2 \frac{S_0(0)}{a_k}
- \tshalf \, L^2 \sum_{i\neq k} \frac{S_i(a_i) }{a_k-a_i} .
\label{eq:453}
\eea
Although the hyperelliptic approximation (\ref{eq:453})
contains one-instanton ($\O(L^2)$) contributions to $a_{D,k}$
it cannot be the complete one-instanton result, 
because of the presence of  unacceptable
$L^2 \log a_k$  and $L^2 \log (a_k+a_j)$ type terms.

\noindent {\bf (b) Corrections to the hyperelliptic approximation}

{}From the results of appendix C, eq.~(\ref{eq:B4}),
the $\O(L^2)$ correction to the hyperelliptic approximation is 
\bea
(2\pi i a_{D,k})_\corr 
& = & \oint_{B_{k}} \l_\corr
			 \nonumber\\ [.1in]
& = & - 2L^2 \int^{x^-_k}_{x^-_1} dx \: {C(x) \over \g^2(x) } 
		\nonumber\\ [.1in]
& =  & - L^2 \int^{x^-_k}_{x^-_1} dx \frac{\prod_i (x-e_i)}
{x^2\prod_i(x + e_i)^2} 
		\nonumber \\ [.1in]
&=  & - \tsquarter (-1)^N L^2 \int^{x^-_k}_{x^-_1} dx \frac{\bg (x)} {\bC^2(x)} \; .
\label{eq:51}
\eea
where 
$\bC(x)$ and $\bg(x)$ are obtained from $C(x)$ and $\g(x)$ respectively
by letting  $ e_i \to -e_i$. 
\bea
\bC(x) & = & \tshalf x^2 \prod^N_{i=1} (x+e_i) \nonumber \\ [.1in]
\bg(x) & = & (-1)^N \, x^2 \prod^N_{i=1} (x-e_i) \; . 
\label{eq:52}
\eea
We expand the integrand of (\ref{eq:51}) in partial fractions
\be
\frac{\bg (x)}{\bC^2(x)}  = 
\sum^{2}_{p=1} 
\frac{\bQ^{(2,0,0)}_{0,p}}{x^p}
+ \sum^N_{j=1} \sum^{2}_{p=1} 
\frac{\bQ^{(2,0,0)}_{j,p}} {(x+e_j)^p}
\label{eq:53}
\ee
which results in
\be
(2\pi i a_{D,k})_\corr 
  =  - \tsquarter (-1)^N L^2
\left[
\bQ^{(2,0,0)}_{0,1} \log e_k 
+ 
\sum^N_{j=1} \bQ^{(2,0,0)}_{j,1} \log (e_k+e_j)
 - \frac{\bQ^{(2,0,0)}_{0,2}}{e_k} 
 - \sum^N_{j=1} \frac{\bQ^{(2,0,0)}_{j,2}}{e_k+e_j }
\right] \; . 
\label{eq:54}
\ee
As before, we introduce residue functions $\bS_j(x)$ and $\bS_0(x)$,
defined  by
\bea
\frac{\bg(x)}{\bC^2(x)} &=& \frac{\bS_j(x)}{(x+e_j)^2} = \frac{\bS_0(x)}{x^2} 
\label{eq:54b}
\eea
in terms of which the partial fraction coefficients may be expressed
\bea
\bQ^{(2,0,0)}_{0,1} 
& = &
\frac{\pa \bS_0}{\pa x} (0) 
	\nonumber \\ [.1in]
\bQ^{(2,0,0)}_{j,1} 
& = &
 \frac{\pa \bS_j}{\pa x} (-e_j) 
	\nonumber \\ [.1in]
\bQ^{(2,0,0)}_{0,2} 
&=& 
\bS_0(0)  
	\nonumber \\ [.1in]
\bQ^{(2,0,0)}_{j,2} 
& = & 
\bS_j(-e_j)        .
\label{eq:54c}
\eea
{}From the explicit expressions for the residue functions
\bea
\bS_0(x) &=&
\frac{4(-1)^N \prod^N_{i=1} (x-e_i) }
{\prod_i (x+e_i)^2}
\nonumber\\ [.1in]
\bS_j(x) &=& \frac{4(-1)^N\prod^N_{i=1} (x-e_i)} 
{x^2\prod_{i\neq j} (x+e_i)^2}
\label{eq:55}
\eea
and eq. (\ref{eq:34b}), we ascertain
\bea
\frac{\pa\bS_0}{\pa x} (0)
&=&  -(-1)^N \frac{\pa S_0}{\pa x} (0) 
	\nonumber \\ [.1in]
\frac{\pa\bS_j}{\pa x} (-e_j)
& = & -(-1)^N \frac{\pa S_j}{\pa x} (e_j)
	\nonumber \\  [.1in]
\bS_0 (0) 
& = & (-1)^N S_0 (0)
	\nonumber\\ [.1in]
\bS_j (-e_j) & = & (-1)^N S_j (e_j)
\label{eq:515}
\eea
Combining (\ref{eq:54}) with (\ref{eq:54c}) and (\ref{eq:515}),
we obtain the contribution  to the dual order parameters from
the correction $\l_\corr$ to the hyperelliptic SW differential, 
accurate to one-instanton order.  
Since the entire correction is $\O(L^2)$,
we may replace $e_i$ by $a_i$ throughout,
resulting in 
\be
(2\pi i a_{D,k})_\corr
 =  \tsquarter L^2   \left[
	 \frac{\pa S_0}{\pa x} (0) \log  a_k  
+ \sum^N_{j=1} \frac{\pa S_j}{\pa x} (a_j) \log  (a_k + a_j)
+  \frac{S_0(0)}{a_k} 
+ \sum^N_{j=1} \frac{S_j(a_j)}{a_k+a_j} 
\right] \;. 
\label{eq:516}
\ee

\noindent{\bf 6. ~The Prepotential}
\renewcommand{\theequation}{6.\arabic{equation}}
\setcounter{equation}{0}

Combining the results (\ref{eq:453}) and (\ref{eq:516})
of the preceding section, 
one obtains the following expression for the dual order parameters,
accurate to the one-instanton level:
\bea
2 \pi i a_{D,k}
& = &
[N + 2 + (N+2) \log (-1) + 2 \log L] a_k  
				\nonumber \\ [.2 in]
&&-  2 \sum_{i\neq k} (a_k-a_i) \log (a_k-a_i) +
      \sum_i (a_k+a_i)  \log (a_k+a_i) -  2 a_k \log a_k 
				\nonumber \\ [.2 in]
&&+ L^2 \left[ \quarter \frac{\pa S_k}{\pa x} (a_k) \;
   	+ {\half}  \frac{S_0(0)}{a_k}
	- \half \, \sum_{i\neq k} \frac{S_i(a_i) }{a_k-a_i} 
	+ {\quarter}  \sum^N_{j=1} \frac{S_j(a_j)}{a_k+a_j} \right]
				\nonumber \\ [.2in]
&&  - ( k \to 1 )
\label{eq:61}
\eea
where we have restored the dependence on the lower integration
limit that was suppressed throughout the calculation,
and where
\bea
S_k(x) &=& \frac{4(-1)^N\prod^N_{i=1} (x+a_i)} 
{x^2\prod_{i\neq k} (x-a_i)^2}
\nonumber\\ [.1in]
S_0(x) &=&
\frac{4(-1)^N \prod^N_{i=1} (x+a_i) }
{\prod_i (x-a_i)^2} \; .
\label{eq:62}
\eea
Note that since the residue functions 
$S_k(x)$ and $S_0(x)$ appear in (\ref{eq:61}) multiplied by $L^2$,
we have replaced 
the unrenormalized order parameters $e_i$ 
in the original definitions (\ref{eq:34b})
with renormalized order parameters $a_i$
in (\ref{eq:62}).

The prepotential $\F(a)$ is found via (\ref{eq:25b})
when written in terms of the independent variables
$a_2, \ldots, a_k$.
Since, using (\ref{eq:C4}) and (\ref{eq:C5}),
$a_1$ obeys the constraint 
\be 
\sum_{j=1}^N a_j = 0
\label{eq:62b}
\ee
for a massless hypermultiplet,
if $\F(a)$ is written in terms of all the variables $a_j$,
(\ref{eq:25b}) becomes
\be
a_{D,k}  =  \frac{\pa \F}{\pa a_k}  -\frac{\pa \F}{\pa a_1}.
\ee
To one-instanton order this becomes
\bea
a_{D,k} = {\pa \over \pa a_k}  
\left[ \F_{\cl} +  \F_{\oneloop} 
+ \L^{N+2} \F_{\oneinst} + \cdots \right] - (k \to 1).
\label{eq:63}
\eea
Integrating (\ref{eq:61}), we obtain 
\bea
 \F_{\cl} + \F_{\oneloop} 
&=& \frac{1}{4 \pi i} 
[ {\textstyle{ 3 \over 2}} (N+2) + (N+2)   \log (-1)  + 2 \log 2 ] \sum_j a_j^2
 \label{eq:610} \\ [.2 in]
 && +  {i\over 8 \pi} 
\left[ \sum^N_{i,j=1} (a_i-a_j)^2 \log {(a_i-a_j)^2\over \L^2}  
 - \sum_{i<j} (a_i + a_j)^2 \log {(a_i+a_j)^2\over \L^2} \right]
 \nonumber
\eea
and
\be
\F_{\oneinst} 
 = {1\over 2\pi i}
 \left[ - \frac{1}{2} \, S_0(0) +  \frac{1}{4} \, \sum_k S_k (a_k) \right]
\label{eq:611}
\ee
a beautiful, concise result in view of the lengthy calculations required.
To arrive at (\ref{eq:611}), we employed the identities
\bea
\frac{\pa}{\pa a_k} [ S_k (a_k)] 
&=& 
\frac{\pa S_k}{\pa x} (a_k)   + \frac{S_k(a_k)}{2a_k} 
	\nonumber \\[.2in]
\frac{\pa}{\pa a_k} \left[ \sum_{i \neq k} S_i (a_i) \right]
& = &
- 2 \sum_{i \neq k} \frac{S_i(a_i)}{a_k-a_i} 
+  \sum_{i \neq k} \frac{S_i(a_i)}{a_k+a_i} 
	\nonumber  \\[.2in]
 \frac{\pa}{\pa a_k} [S_0(0)] 
& = & - \frac{S_0(0)}{a_k} 
\label{eq:67}
\eea
which follow directly from (\ref{eq:62}).
Note that the results (\ref{eq:610}) and (\ref{eq:611}) 
are invariant under permutation of the $a_k$, 
and hence the Weyl group of SU(N).

The calculation above is for a massless hypermultiplet
in the antisymmetric representation. 
Shifting
\be
a_i \longrightarrow a_i + \frac{m}{2}
\label{eq:68}
\ee
in (\ref{eq:610}) and (\ref{eq:611}) gives
a result for a hypermultiplet with mass $m$
that is consistent with the known cases (see next section).

\noindent{\bf 7. ~ Tests of the One-Instanton Predictions}
\renewcommand{\theequation}{7.\arabic{equation}}
\setcounter{equation}{0}

Comparison of (\ref{eq:24b}) with (\ref{eq:610})  shows that the 
Landsteiner-Lopez curve
correctly predicts the perturbative one-loop prepotential.  
Different curves, however,
can provide the same predictions to one-loop order, 
$G_2$ \cite{031} and $E_6$ \cite{032} being well-known examples.
Therefore, one needs to compare (at least) the one-instanton
predictions of the curve with field theoretic results
before one can be certain that the curve correctly describes the 
low-energy field theory. 
One should hold M-theory to this standard.

The one-instanton contribution to the prepotential for 
a hypermultiplet in the antisymmetric representation
is already known for SU(2), SU(3), and SU(4)
from other considerations.
The antisymmetric representation of SU(2) is the trivial representation
so the case of SU(2) corresponds to pure SU(2) gauge theory. 
We compare (\ref{eq:611}) for $N=2$
with the results of DKP \cite{022},
eq. (4.33b) for $N_c=2$ and $N_f=0$,
finding agreement with the change of scale
$L^2 = {1\over 16} \bar\L^2_{\rm DKP}$.
 
The antisymmetric representation of SU(3) should give the same result
as the defining representation.
Comparing (\ref{eq:611}) (with the shift (\ref{eq:68})) for $N=3$
with the results of
DKP \cite{022}, eq. (4.33b),
for $N_c=3$ and $N_f=1$,
and using $a_1+a_2+a_3 = 0$,
we find agreement, again with a change of the quantum scale.

The antisymmetric representation of SU(4) is equivalent to the 
the defining representation of SO(6).
In particular, the weights of the 
antisymmetric representation of SU(4) are
\be
\pm (a_1 + a_2),  \; \pm (a_2+a_3), \; \pm (a_3+a_1) \; .
\label{eq:74}
\ee
The weights of the fundamental representation of SO(6) are
\be
\pm d_i \hspace{.4in} (i=1,2,3) \; .
\label{eq:75}
\ee
These weights are identified as follows:
\bea
d_1 & = & a_1 + a_2 \nonumber \\
d_2 & = & a_2 + a_3 \\
d_3 & = & a_3 + a_1 \; . \nonumber 
\label{eq:76}
\eea
DKP \cite{023} (and also ref.~\cite{028})
give the one-instanton contribution for 
SO(6) with one massless hypermultiplet in the defining representation 
\be
\F_{\oneinst} = \frac{1}{4\pi i} \, \L_{\rm DKP}^6 
\sum^3_{k=1} \Sigma_k (d_k)
\label{eq:72}
\ee
where
\be
\Sigma_k(x) = \frac{x^6}{(x+d_k)^2 \prod_{j\neq k} (x^2 - d^2_j)^2} \; .
\label{eq:73}
\ee
Using $a_4 \equiv -(a_1+a_2+a_3)$,  
we find that (\ref{eq:611}) for $N=4$ is equivalent to (\ref{eq:72}), 
with a change of the quantum scale.  
Thus, we find agreement for this case as well.

\noindent{\bf 8. ~Concluding Remarks}

In this paper we derived the one-instanton contribution to the
prepotential for $N$=2 supersymmetric  SU(N) gauge theory 
with a matter hypermultiplet in the antisymmetric approximation, 
using the non-hyperelliptic curve obtained from M-theory by 
Landsteiner and Lopez.  
To carry out this calculation, 
we developed a systematic perturbation theory for the curve
and Seiberg-Witten differential,
where the zeroth order curve is hyperelliptic.  
Our results for the Landsteiner-Lopez curve 
agree with known results for SU(2), SU(3), and SU(4), 
and provide predictions for SU($N$), $N\geq$5,
which could be checked against future ``microscopic" instanton
calculations \cite{030} in $N$=2 supersymmetric gauge theories.

A companion \cite{029} to this paper describes a similar
calculation for the symmetric representation of SU(N).  
It is extremely important to continue 
to develop the bridge between string
theory and field theory, and in particular to verify the predictions
of geometric engineering and M-theory for $N$=2 supersymmetric
gauge theories, so as to gain confidence in the validity of string
theory predictions of field theoretic phenomena.  We believe that
the methods of this paper will prove useful for that purpose.

{\bf Acknowledgements}

We wish to thank 
Isabel Ennes, 
Jos\'{e} Isidro, 
Michael Mattis,
Jo\~{a}o Nunes,
and 
\"Ozg\"ur Sar{\i}o\~{g}lu
for discussions on various aspects of this work.  
HJS wishes to thank the Physics Department of Harvard
University for their hospitality during the spring semester of 
1998.

Finally, we wish to express our appreciation for the beautiful
work of D'Hoker, Krichever, and Phong, which greatly influenced
this paper.

\newpage

\noindent{\large\bf Appendix A: Hyperelliptic Perturbation Theory}
\renewcommand{\theequation}{A.\arabic{equation}}
\setcounter{equation}{0}

Consider the cubic curve
\be
y^3 + 2A (x) y^2 + B(x) y + \ell(x) =0 \;.
\label{eq:A3}
\ee
We may eliminate the quadratic term by changing variables
\be
w = y + \tstwothirds \, A \; ,
\label{eq:A4}
\ee
yielding
\be
w^3 + \left(B - {\textstyle {4\over3}} A^2\right) w 
+ \left({\textstyle{16\over 27}} A^3 - {\tstwothirds} A B  + \ell \right) = 0.
\label{eq:A5}
\ee
The solutions of (\ref{eq:A5}) satisfy
\bea
w_1 + w_2 + w_3 
&=& 0  \nonumber \\ [.1in]
w_1 w_2 + w_1 w_3 + w_2 w_3 
& = & B - {\textstyle {4\over3}} A^2    \nonumber \\ [.1in]
w_1 w_2 w_3  
& = & -  {\textstyle{16\over 27}} A^3 + {\tstwothirds} A B  - \ell .
\label{eq:A6}
\eea

When $ \ell$ vanishes,  
the curve (\ref{eq:A3}) factors into $y=0$ and a hyperelliptic curve
$y^2 + 2A(x) y + B(x)  = 0$.
We will use the solutions of this as a 
starting point for a perturbative expansion in $\ell$.
We remark that though $\ell$ is proportional to (some power of) $\L$,
our perturbative expansion is {\it not} equivalent to an expansion
in $\L$.
In particular, since $A(x)$ and $B(x)$ have $\L$ dependence,
the hyperelliptic approximation, 
which is zeroth order in $\ell$,
contains all orders in $\L$.

For $\ell = 0 $, the solutions of (\ref{eq:A3}) are
\bea
\by_1 & = & - A - r \nonumber \\
\by_2 & = & - A + r \nonumber \\
\by_3 & = & 0
\label{eq:A7}
\eea
where $r = \sqrt{A^2 - B}$. 
Equivalently, the solutions of (\ref{eq:A5}) are
\bea
\bw_1 &  = & - \tsthird A - r \nonumber \\
\bw_2 &  = & - \tsthird A + r \nonumber \\
\bw_3      & = & \tstwothirds A \; .
\label{eq:A8}
\eea

When $\ell \neq 0$, 
the solutions to (\ref{eq:A5})
can be written as expansions around the hyperelliptic solutions
(\ref{eq:A8}),
\be
w_i = \bw_i + \d w_i = \bw_i + \a_i \ell + \b_i \ell^2 + \cdots
\label{eq:A9}
\ee
Substituting this into (\ref{eq:A6}), we obtain
\bea
\d (w_1 + w_2 + w_3) & = & 0 \nonumber \\
\d (w_1w_2 + w_2 w_3+ w_3w_1) & = & 0 \nonumber \\
\d (w_1  w_2  w_3) & = & -\ell     
\label{eq:A12}
\eea
To first order in $\ell$, we have
\bea
\a_1 + \a_2 + \a_3 & = & 0 \nonumber \\
\bw_1 \a_1 + \bw_2 \a_2 + \bw_3 \a_3  & = & 0 \nonumber\\
\bw_2 \bw_3 \a_1 + \bw_1 \bw_3 \a_2 + \bw_1 \bw_2 \a_3  & = & -1 \ .
\label{eq:A13}
\eea
which can be solved to give
\bea
\a_1 & = & \frac{1}{(\bw_1-\bw_2)(\bw_3-\bw_1)}  
~=~ -  {1 \over 2r(A+r)} ~=~ - \, {A-r \over 2Br}
\nonumber \\[.1in]
\a_2& = & \frac{1}{(\bw_2-\bw_3)(\bw_1-\bw_2)}  
~=~ {1 \over 2r(A-r)} ~=~ {A+r \over 2Br}
\nonumber \\[.1in]
\a_3& = & \frac{1}{(\bw_3-\bw_1)(\bw_2-\bw_3)}  
=  - {1 \over B} \; . 
\label{eq:A14}
\eea
The solutions of (\ref{eq:A3}) are therefore
\bea
y_1 & = & - A - r  - \ell {A-r \over 2Br} + {\O} (\ell^2 )\nonumber \\
y_2 & = & - A + r  + \ell {A+r \over 2Br} + {\O} (\ell^2 )\nonumber \\
y_3 & = &   - \ell {1 \over B} + {\O} (\ell^2 )
\label{eq:A15}
\eea
Note that, to this order, $y_3$ does not exhibit a branch-cut,
while sheets labelled by $y_1$ and $y_2$ are linked by the branch-cuts
that arise from $r = \sqrt{A^2 -B}$.

The second order corrections in (\ref{eq:A9}) must obey
\bea
\b_1 + \b_2 + \b_3 
& = & 0 \nonumber \\
\bw_1  \b_1 + \bw_2  \b_2 + \bw_3  \b_3 
& = & \a_1\a_2 + \a_2\a_3 + \a_3\a_1 \nonumber\\
\bw_2\bw_3 \b_1  +  \bw_3 \bw_1 \b_2 + \bw_1 \bw_2 \b_3  
& = & 
- \a_1\a_2 \bw_3 - \a_2\a_3 \bw_1 - \a_3\a_1\bw_2 
\label{eq:A18}
\eea
which have the solution
\bea
\b_1 & = & \frac{3 \bw_1}{(\bw_1-\bw_2)^3(\bw_3-\bw_1)^3}  \nonumber \\[.1in]
\b_2 & = & \frac{3 \bw_2}{(\bw_2-\bw_3)^3(\bw_1-\bw_2)^3}  \nonumber \\[.1in]
\b_3 & = & \frac{3 \bw_3}{(\bw_3-\bw_1)^3(\bw_2-\bw_3)^3}  \; . 
\label{eq:A19}
\eea
In particular, this implies
\be
y_3  =  - \ell  {1 \over B} - \ell^2 \frac{2A}{B^3} + {\O} (\ell^3)
\label{eq:A20}
\ee
so that branch cuts in $y_3$ do not appear to second order in $\ell$.  

We have verified that the $\O (\ell^2)$ terms of (\ref{eq:A15}) will
not contribute to the one-instanton correction to the prepotential;
the first order solutions suffice.

\newpage

\noindent{\large\bf Appendix B: The Landsteiner-Lopez Curve}
\renewcommand{\theequation}{B.\arabic{equation}}
\setcounter{equation}{0}

The spectral curve proposed by Landsteiner and Lopez  \cite{027} 
for a hypermultiplet in the antisymmetric representation is
\be
y^3  + \left[ f(x)  + 3\L^{N+2} \right] y^2 
 +  \L^{N+2} \left[ g(x)  + 3\L^{N+2} \right] y 
	+ \L^{3N+6} = 0 \; , 
\label{eq:A1}
\ee
where
\bea
f(x) & = & x^2 \prod^N_{i=1} (x-e_i) \nonumber \\
g(x) & = & (-1)^N \: x^2 \prod^N_{i=1} (x+e_i)
\label{eq:A23}
\eea
Since $f(x) = g(-x)$, the curve (\ref{eq:A1}) is invariant under
the involution \cite{027}
\bea
\left\{
\begin{array}{lcl}
y & \rightarrow & \L^{2N+4} /y \\
x & \rightarrow & -x
\end{array} \right. \; \; .
\label{eq:A2}
\eea
Consequently,  if $y(x)$ is a solution of (\ref{eq:A1}), then
$\tilde{y} (x) \equiv \L^{2N+4}/y(-x)$ is also a  solution.

Solutions of (\ref{eq:A1}) in the hyperelliptic expansion
introduced in Appendix \AppA~are 
\bea
y_1(x) & = & - A - r  - {L^6 (A-r) \over 2Br} + \cdots\nonumber \\
y_2(x) & = & - A + r  + {L^6 (A+r) \over 2Br} + \cdots   \nonumber \\
y_3(x) & = &   - {L^6 \over B} + \cdots 
\label{eq:A15b}
\eea
where
\bea
L^2 & = & \L^{N+2} \nonumber \\
A & = & \tshalf f(x) + {\textstyle {3\over 2}}  L^2 \nonumber \\
B & = & L^2 g(x) + 3L^4\nonumber\\
r & = & \sqrt{A^2 -B}
\label{eq:A22}
\eea
It may be verified to the order we are working  
that the involution (\ref{eq:A2}) permutes these solutions
as follows:
\bea
\tilde{y}_1(x) & = & y_3(x) \nonumber \\
\tilde{y}_2(x) & = & y_2(x) \nonumber \\ 
\tilde{y}_3(x) & = & y_1(x) 
\label{eq:A26}
\eea
where
$\tilde{y}_i(x) \equiv L^4 / y_i (-x)$.

{}From  (\ref{eq:A26}), 
as well as explicit analysis, 
one deduces the following structure for the three-fold
branched covering of the sphere.

1) Sheets corresponding to $y_1$ and $y_2$ are connected by $N$
square-root branch-cuts centered about $x=e_i \;\;\; (i=1$  to $N)$.

2) Sheets corresponding to $y_2$ and $y_3$ are connected by $N$
square-root branch-cuts centered about $x=-e_i\;\;\; (i=1$  to $N)$.

3) From  (\ref{eq:A1}) and (\ref{eq:A23}), 
sheets $y_1,\;y_2,\; y_3$ coincide at $x=0$, where
\be
y_{1,2,3} = - L^2 + {\O} (x) \; . 
\label{eq:A27}
\ee
There are, however, no branch-cuts at $x=0$.

This three-sheeted branched covering of the sphere 
is a Riemann surface of genus $2N-2$.
{}From (\ref{eq:A27}),
one sees that the hyperelliptic perturbation theory 
breaks down for $x \ll  L$.  
This does not, however, 
change the dual order parameters $a_{D,k}$ 
or, therefore, the prepotential.

\newpage

\noindent{\large\bf Appendix C: The Seiberg-Witten Differential}
\renewcommand{\theequation}{C.\arabic{equation}}
\setcounter{equation}{0}

The Seiberg--Witten (SW) differential for the curve  (\ref{eq:26})  
is 
\be
\l = x \frac{dy}{y}
\label{eq:B1}
\ee
which takes a different value on each of the sheets labeled by
the solutions  (\ref{eq:29}). 
To the order we are working, 
sheet 3 is disconnected from sheets 1 and 2, 
so we only consider $y_1$ and $y_2$.  
We write these solutions in the hyperelliptic expansion
(\ref{eq:29}) as
\bea
y_1 
&=& - (A+r) \left[ 1 + \frac{L^6 (A-r)}{2Br (A+r) } + \ldots  \right]
\nonumber\\
&=& - (A+r) \left[ 1 + \frac{L^6 (2A^2 - B -2Ar)}{2B^2 r } + \ldots  \right]
\eea
with $y_2$ obtained from this by letting $r \rightarrow -r$.
The SW differential on sheet 1 is
\be
\l_1   =  x \frac{dy_1}{y_1}
=   {x\over A+r} d(A + r)
+ x ~ d \left[ \frac{L^6 (2A^2 - B -2Ar) }{2B^2 r }\right] 
+ \ldots  
\ee
where the first term is 
\bea
\l_\hea 
&=&  {x\over A+r} d(A + r)
\nonumber \\[.1in]
&\simeq&   {x\over r} \left( A^\prime - {A B^\prime \over 2B} \right) dx
\nonumber \\[.1in]
&=&   {x \left( {A^\prime\over A} - {B^\prime \over 2B} \right) \over 
\sqrt{1 - {B\over A^2}}} dx
\label{eq:B2}
\eea
the usual hyperelliptic form of  the
SW differential,
while the second term is the correction to the differential
\bea
\l_\corr 
&=& x ~  d \left[ \frac{L^6 (2A^2 - B -2Ar) }{2B^2 r }\right] 
 \nonumber \\[.1in]
&\simeq & -~ 
\left[ \frac{L^6 (2A^2 - B) }{2B^2 r } \right] dx \nonumber \\[.1in]
& =  & -~ 
 { L^6 \left(A - {B\over 2A}  \right)  \over
	 B^2 \sqrt{1 - {B\over A^2}} } dx
\label{eq:B3}
\eea
where $\simeq$ in both (\ref{eq:B2}) and (\ref{eq:B3})
means we have dropped terms that
do not contribute  to period integrals around the $A$ or $B$ cycles.
The SW differential on the second sheet, $\l_2$,
is obtained by taking the negative sign for the square root
in both (\ref{eq:B2}) and (\ref{eq:B3}). 

For the Landsteiner-Lopez curve,
we have (\ref{eq:27a})
\bea
A(x) & = & C(x) + \O (L^2) 
\nonumber \\ 
B(x) & = & L^2 \g(x) +\O(L^4)
\eea
so the $\O(L^2)$ correction to the SW differential becomes simply
\be
\l_\corr = - L^2 {C \over \g^2}~dx
\label{eq:B4}
\ee

\noindent{\large\bf Appendix D: Identities}
\renewcommand{\theequation}{D.\arabic{equation}}
\setcounter{equation}{0}

In this Appendix we derive three identities used in the paper,
namely
\be
\sum_{k=1}^N e_k \, R_k(e_k) + \frac{3}{4} \,
S_ 0 (0) = 0
\label{eq:C1}
\ee 
\be
\sum_{k=1}^N R_k(e_k) + \frac{1}{4} \,
\frac{\pa S_0}{\pa x} (0) = 0 
\label{eq:C2} 
\ee
and
\be
\sum^N_{k=1} \left[
R_k (e_k)  
- \frac{1}{4} \frac{\pa S_k}{\pa x} (e_k) 
\right] = 0.
\label{eq:C3}
\ee
These identities allow us to eliminate all reference to the residue function
$R_k$ in the final expressions for the prepotential.
Also recall that by (\ref{eq:311})
\be
a_k = e_k + L^2 \left[ \frac{1}{4} \frac{\pa S_k}{\pa x} (e_k) - R_k (e_k) \right] + \ldots
\label{eq:C4}
\ee
so (\ref{eq:C3}) 
implies that
\be
\sum_{k=1}^N \, a_k = \sum_{k=1}^N \, e_k
\label{eq:C5}
\ee
to $\O(L^2)$.

To prove the identities above,
we define
\be
F(x) = \frac{3}{x \prod^N_{i=1} (x-e_i)} \; .
\label{eq:C20} 
\ee 
Then
\be
(x-e_k) F(x) = \frac{3}{x \prod _{i\neq k} (x-e_i)} \; .
\label{eq:C21} 
\ee 
and 
\be
\sum^N_{k=1} 
[(x-e_k) F(x)]_{x=e_k}  = \sum_k \frac{3}{e_k \prod _{i\neq k} (e_k-e_i)} 
 =  \sum_k \, e_k R_k (e_k) \; .
\label{eq:C22} 
\ee
Also
\be
 [x \, F(x) ]_{x=0}  =  \frac{3(-1)^N}{\prod^N_{i=1} e_i} 
  =  \frac{3}{4} \, S_0(0) \; .
\label{eq:C23} 
\ee
Thus
\be
\sum_k \, e_k R_k (e_k) + \frac{3}{4} \, S_0(0)
 = \sum^N_{k=1} [(x-e_k) F(x)]_{x=e_k} + [x\, F(x)]_{x=0}  
\label{eq:C24}
\ee
The right hand side is the sum of the residues of $F(x)$.
This vanishes,  since $F(x)$ has no poles at infinity,
thus proving (\ref{eq:C1}).

Next define
\be
H(x)  = \frac{3}{x^2 \prod_{i=1}^N (x-e_i)} \; .
\label{eq:C25}
\ee
The sum of residues at $x=e_k$ is
\be
\sum^N_{k=1} [(x-e_k) H(x)]_{x=e_k}
= \sum^N_{k=1} R_k(e_k)  
\label{eq:C26}
\ee
$H(x)$ also has a double pole at $x$=0, with residue 
\bea
\left. {\pa \over \pa x} \: \frac{3}{\prod_i (x-e_i)} \right|_{x=0} & = & 
- \sum_k \:\left. \frac{3}{(x-e_k)\prod_i (x-e_i)} \right|_{x=0} \nonumber \\
& = &  (-1)^N \sum_k \:\frac{3}{e_k \prod_i e_i } \nonumber\\
& = &  \frac{3}{4} \: S_0(0) \sum_k \, \frac{1}{e_k} \; .\nonumber \\
& = &  \frac{1}{4} {\pa S_0 \over \pa x} (0) 
\label{eq:C27}
\eea
where the last step follows directly from (\ref{eq:34b})
Thus, the vanishing of the sum of residues of $H(x)$ 
implies the identity (\ref{eq:C2}).

Finally, define
\be
K(x) = \frac{\g(x)}{C^2(x)} =
\frac{4(-1)^N \prod^N_{i=1} (x+e_i)}
{x^2  \prod^N_{i=1} (x-e_i)^2} \; .
\label{eq:C30}
\ee
Then residues of $K(x)$ are given by
\bea
{\rm Res} \: K(x)|_{x=e_k} 
&=& \left[ \frac{\pa}{\pa x} 
S_k(x) \right]_{x=e_k}
\nonumber \\[.1in]
{\rm Res} \: K(x)|_{x=0}
& =&  \left[ \frac{\pa}{\pa x} 
S_0(x) \right]_{x=0} \; .
\label{eq:C31}
\eea
The vanishing of the sum of residues of $K(x)$ implies
\be
\frac{\pa S_0}{\pa x} (0) + 
\sum_k \frac{\pa S_k}{\pa x} (e_k) 
=0
\label{eq:C29}
\ee
which together with (\ref{eq:C2}) implies 
the identity (\ref{eq:C3}).

\end{document}